\documentclass[twocolumn,prd,nofootinbib,superscriptaddress,eqsecnum,tightenlines,8pt]{revtex4}

\usepackage{fancyhdr}
\usepackage{amsmath,amsfonts,amssymb}
\usepackage[colorlinks,citecolor=blue,linkcolor=blue,urlcolor=blue]{hyperref}
\usepackage[english]{babel}

\usepackage{color}
\definecolor{red}{rgb}{1,0,0}
\definecolor{gre}{rgb}{0,0.6,0}
\definecolor{blu}{rgb}{0,0,1}

\usepackage{mathenv}
\usepackage{ulem}

\usepackage[pdftex]{graphicx}
\usepackage{mathrsfs} 
\DeclareGraphicsExtensions{.jpg, .JPG , .png , .pdf, .gif}
\def\be{\begin{equation}}
\def\ee{\end{equation}}

\begin{document}

\title{A status report on the phenomenology of black holes in loop quantum gravity:\\ Evaporation, tunneling to white holes, dark matter and gravitational waves}

\date{\today}

\author{Aur\'elien Barrau}
\email{Aurelien.Barrau@cern.ch}
\affiliation{
Laboratoire de Physique Subatomique et de Cosmologie, Universit\'e Grenoble-Alpes, CNRS-IN2P3\\
53, Avenue des Martyrs, 38026 Grenoble cedex, France\\
}%

\author{Killian Martineau}
\email{martineau@lpsc.in2p3.fr}
\affiliation{
Laboratoire de Physique Subatomique et de Cosmologie, Universit\'e Grenoble-Alpes, CNRS-IN2P3\\
53, Avenue des Martyrs, 38026 Grenoble cedex, France\\
}%

\author{Flora Moulin}
\email{moulin@lpsc.in2p3.fr}
\affiliation{
Laboratoire de Physique Subatomique et de Cosmologie, Universit\'e Grenoble-Alpes, CNRS-IN2P3\\
53, Avenue des Martyrs, 38026 Grenoble cedex, France\\
}%

\begin{abstract}
The understanding of black holes in loop quantum gravity is becoming increasingly accurate. This review focuses on the possible experimental or observational consequences of the underlying spinfoam structure of space-time. It adresses both the aspects associated with the Hawking evaporation and the ones due to the possible existence of a bounce. Finally, consequences for dark matter and gravitational waves are considered.
\end{abstract}

\maketitle

\section{Introduction}

The Planck length is $10^{15}$ times smaller than scales probed at colliders. Linking quantum gravity with observations is therefore extremely hard (see, {\it e.g.}, \cite{Barrau:2017tcd} for a recent review and \cite{Hossenfelder:2009nu,Liberati:2011bp,AmelinoCamelia:2008qg} for complementary viewpoints). Most works devoted to the connection of quantum gravity with experiments are focused on cosmology or astroparticles physics. In the cosmological sector, the main goal consists in calculating scalar and tensor power spectra (see, {\it e.g.}, \cite{eucl3,Agullo2}), together with the background dynamics (see, {\it e.g.}, \cite{Martineau:2017sti,Ashtekar:2011rm}). In the astroparticle physics sector, the main idea is to investigate the possible consequences of the granular structure of space (see, {\it e.g.}, \cite{Vasileiou:2015wja} for a recent investigation). \\

Although black holes (BH) have been intensively studied in quantum gravity, those investigations were mostly disconnected from observations and focused on consistency issues. Recovering, at the leading order, the Bekenstein-Hawking entropy is, for example, obviously a major requirement for all tentative theories (see, {\it e.g.}, \cite{G.:2015sda} and references therein). Curing the central singularity -- understood as a classical pathology -- is another one (see, {\it e.g.}, \cite{Ashtekar:2005qt,Bojowald:2005ah}). Solving the information paradox (see, {\it e.g.}, \cite{Mathur:2009hf} and references therein) would also be highly desirable (this is clearly connected to the previous issues).\\

In this article, we focus on black holes as possible probes for loop quantum gravity (LQG). We begin by a very short summary of the basics of black hole physics in this framework. We then switch to consequences for the Hawking evaporation, considering different possible perspectives. The quite recent (within the LQG setting) hypothesis of black holes bouncing into white holes is presented with the possible associated signals. Finally, we critically review the possible links with dark matter and conclude with prospective for gravitational waves.

\section{Basics of black holes in loop quantum gravity}

The study of black holes is an incredibly fruitful field of theoretical physics. Black holes are simple objects. They are pure geometry. There is no equation of state needed: they are just vacuum solutions to the Einstein equations. This is their first fundamental characteristic. The second specificity of black holes lies in the fact that they are (classically) scale invariant \cite{Bekenstein:2003dt}. They can, in principle, exist at any mass.\\

As far as quantum gravity is concerned, the major breakthrough came from black holes thermodynamics. Because of the no-hair theorem, in Einstein gravity, the most general stationary black hole geometry is described by the Kerr-Newman (KN) solution with mass $M$, electric charge $q$ and angular momentum $j$ as the only parameters.
One can define three length scales characterizing the BH \cite{Bekenstein:2003dt}: $m\equiv GMc^{-2}$, $Q\equiv\sqrt{G} qc^{-2}$ and $a\equiv jM^{-1}c^{-1}$. There exists a BH solution only when
$Q^{2}+a^{2}\leq m^{2}$. One can show, from the area expression, that
\begin{equation} d(Mc^{2})=\Theta dA+\Phi dQ+\Omega dj
\label{bhlaw}
\end{equation} with
\begin{eqnarray}
\Theta &\equiv& c^{4}(2GA)^{-1}(r_{g}-m),
\\
\Phi &\equiv&q\,r_{g} (r_{g}^2+a^{2})^{-1},
\\
\Omega &\equiv& j\, m^{-1}(r_{g}^2+a^{2})^{-1},
\label{der}
\end{eqnarray}
$r_{g}=2m$ being the gravitational radius. The parameters $\Theta$, $\Phi$ and $\Omega$ can be understood as the surface gravity, the electrostatic potential, and the angular momentum. 

As $Mc^{2}$ is the energy, this equation looks like the
first law of thermodynamics $TdS=dE-\Phi dQ-\Omega dj$. This led to the introduction of a temperature 
\begin{equation} 
T_{H}= (2c\hbar/A)\sqrt{M^{2}-Q^{2}-a^{2}},
\label{TH}
\end{equation}
and entropy
\begin{equation} S_{BH}=A/4\ell_{P}{}^{2},
 \label{S}
\end{equation} 
yealding the evaporation process \cite{Hawking:1974sw}.
The second BH law expresses the fact that the sum of the BH entropy together with the entropy outside the BH cannot decrease. (From now on, except otherwise stated, we use Planck units.)\\

The description of BHs in LQG heavily relies on the concept of isolated horizons (IH) \cite{Ashtekar:1998sp,Ashtekar:1999yj,Ashtekar:2000sz,Lewandowski:1999zs,Lewandowski:2000nh}. This is an intrinsically quasilocal notion which has the advantage of not requiring the knowledge of whole spacetime to determine whether horizons are present, as is the case with event horizons. The most important characteristics of isolated horizons are \cite{G.:2015sda}: their quasilocality, the availability of a Hamiltonian description for the sector of GR containing the IH, the possibility of finding physical versions of the laws of BH thermodynamics and the existence of local definitions of the energy and angular momentum.\\

This article focuses on the consequences and not on the theoretical definition of an LQG BH, but recent pedagogical reviews on BH in LQG can be found, {\it e.g.}, \cite{Perez:2017cmj,Olmedo:2016ddn,Gambini:2013exa,Gambini:2013hna,Roken:2012pt,Agullo:2012zz,DiazPolo:2011np}.

Very schematically, the isolated horizon plays the role of a boundary for the
underlying manifold before quantization. Given the area $A$ of a Schwarzschild BH horizon, the geometry
states of the BH horizon arise from a punctured sphere. Each puncture carries quantum numbers (see, {\it e.g.},
\cite{Ashtekar:1997yu,Ashtekar:2000eq,Corichi:2006wn,Corichi:2006bs} for details): two labels $(j,m)$, where $j$ is a
spin half-integer carrying information about the area and $m$ is the corresponding projection
carrying information about the curvature. They fulfill the condition
\begin{equation}
\label{eq1}
A-\Delta\leq 8\pi \gamma \sum_{p}{\sqrt{j_p(j_p+1)}}\leq A+\Delta,
\end{equation}
where $\gamma$ is the Barbero-Immirzi parameter entering the definition of LQG (see, {\it e.g.}, \cite{Rovelli:2014ssa}), $\Delta$ is the ``smearing'' area parameter (or coarse graining scale) used to recover the classical description and $p$ refers to different punctures. In addition, one requires
\begin{equation}
\label{eq2}
\sum_{p}{m_p=0},
\end{equation}
which means that the horizon has a spherical topology. Many aspects of the BH entropy were studied
in this framework and we shall mention some of them in the following.

\section{Modified Hawking spectrum}

One cannot directly measure the entropy of a BH. So even if some quantum gravity approaches do predict some corrections with respect to the Bekenstein-Hawking law, this can hardly be considered as a smoking gun for observational aspects of quantum geometry. On the other hand, one might observe the evaporation of a black hole. This would require light black holes (the temperature of a solar-mass BH is far below the one of the cosmological microwave background) whose existence is far from obvious. At this stage the Hawking evaporation of BHs therefore remains purely theoretical (although there are some hints that this could have been observed in analog systems \cite{Steinhauer:2015saa}). But it is in principle observable and might constitute a path toward experimental quantum gravity.

\subsection{Global perspective}

The first obvious idea to investigate LQG footprints is to consider the deep planckian regime of an evaporating BH by taking into account the discrete structure of the area operator eigenvalues in LQG. An edge with spin representation $j$ of SU(2) carries an area of eigenvalue 
\begin{equation}
A_j=8\pi \gamma \sqrt{j(j+1)},
\label{eq1}
\end{equation}
where $j$ is, again, a half-integer. A BH surface punctured by $N$ edges therefore exhibits, as explained previoulsy, a spectrum given by 
\begin{equation}
A_j=8 \pi \gamma \sum_{n=1}^N\sqrt{j_n(j_n+1)},
\end{equation}
where the sum is carried out over all intersections of the edges with the isolated horizon. As the area spectrum in discrete, BHs can only make discontinuous jumps and the evaporation spectrum will inevitably be modified.\\

In \cite{Barrau:2011md}, a Monte-Carlo simulation was carried out to investigate to which extent the associated line structure can be discriminated from the usual continuous (enveloppe of the) spectrum. The algorithm was based on an improved version of the method given in \cite{Agullo:2010zz}, enhanced by an efficient numeration scheme based on a breadth-first search. The probability for the transition from a BH state to another is expressed as the exponential of the entropy difference, weighted by the greybody factor. As the optical limit was not satisfactory to derive accurate results, the full greybody factor obtained by solving the wave equation in the (classical) Schwarzschild background was used. The simulation was started at 200 $A_{Pl}$, where $A_{Pl}$ is the Planck area.\\

At each step $n$ of the simulation, starting from a BH mass $M_{n}$, a new mass $M_{n+1}$ is randomly determined within the available spectrum, according to the probability law previously given. A particle type is then randomly selected among the standard model, according to the weighted number of internal degrees of freedom (and among those with a mass smaller than $\Delta M$). The available energy $M_{n}-M_{n+1}$ is assigned to this particle and the process is repeated.  The analysis presented in the following was carried out considering only the emitted photons, that is approximately 1.5\% of the emitted quanta. This choice is motivated by the fact that they keep their initial energy (quarks and gluons lead to jets), they are easy to detect (neutrinos are not), stable (muons or tau leptons do decay) and unaffected by magnetic fields (electrons are).\\

The simulation was repeated many times to account for different possible realizations of the process. As expected the time-integrated spectrum exhibits lines that are not present in the standard Hawking spectrum. The time integrated differential Hawking spectrum scales as $E^{-3}$, where $E$ is the energy of the emitted photons. In this case it becomes a truncated power-law as the available energy is limited. To test to which extent the LQG spectrum can be distinguished from a standard Hawking spectrum, a Kolmogorov-Smirnov (K-S) test was implemented. The K-S statistics measures the distance between the cumulative distribution functions of the considered distributions and can be used for a systematic study of discrimination capabilities.\\

\begin{figure}[!h]
\begin{center}
\includegraphics[scale=0.45]{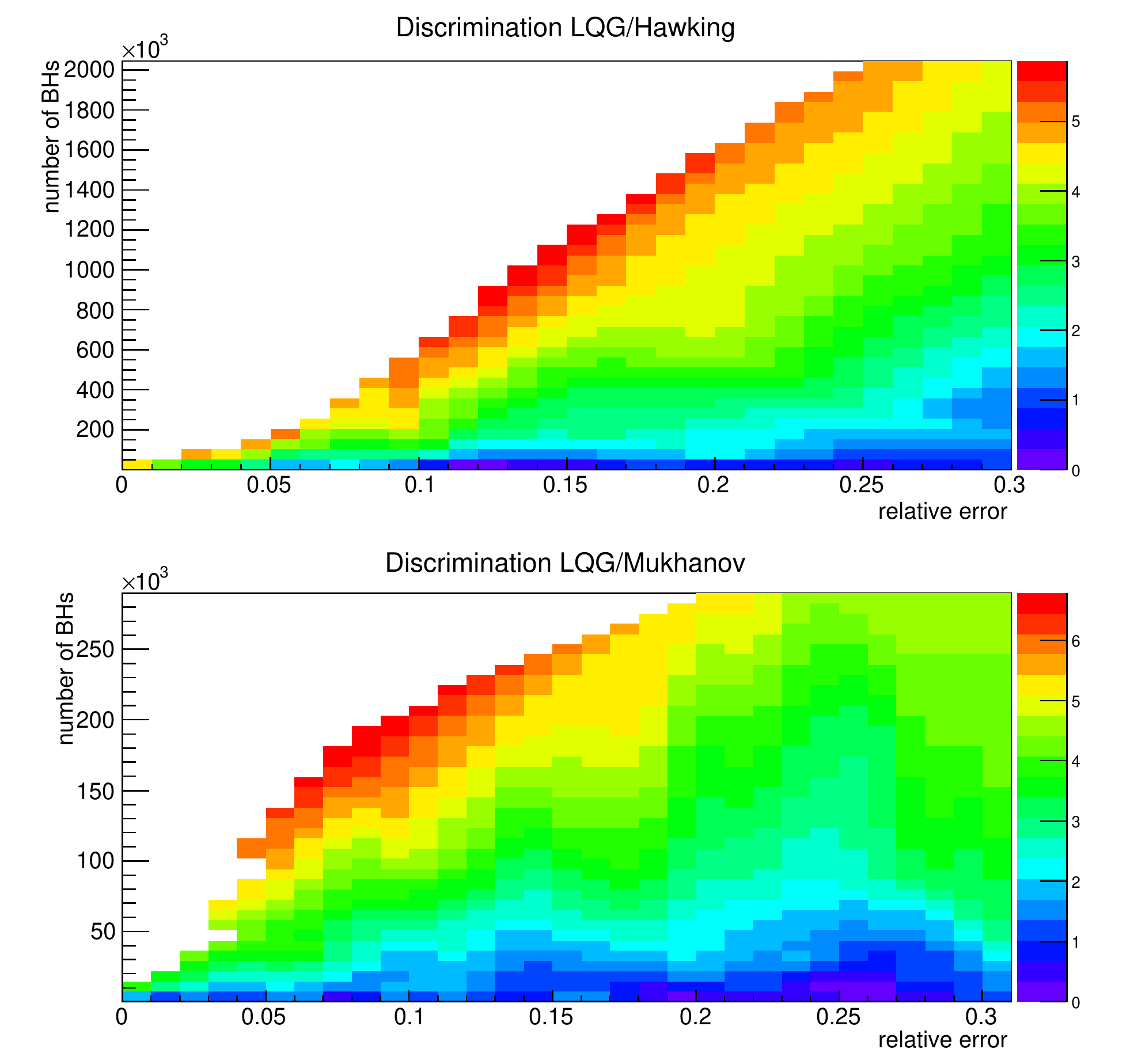}
\caption {Number of BHs that would have to be observed as a function of the relative error on the energy measurement for different confidence levels (the color scale corresponds to the number of standard deviations). Upper plot : discrimination between LQG and the Hawking spectrum. Lower plot : discrimination between LQG and the Mukhanov-Bekenstein hypothesis \cite{Bekenstein:1995ju}. From \cite{Barrau:2011md}.} 
\label{fig0}
\end{center}
\end{figure}

Fig. \ref{fig0} shows the number of evaporating BHs, seen in their final stages, that would be required to discriminate at a given confidence level between the Hawking spectrum and the LQG spectrum, depending on the experimental uncertainty of the measure of the energy of the detected photons. This later parameter is mandatory. If the resolution was infinite a single photon could nearly allow to discriminate, but this is obviously never the case. The results are theoretically appealing but  experimentally challenging.\\ 

Another interesting feature is the following. The end of the evaporation in the LQG framework consists in the emission of a few particles, whose energies are given by the mass difference between BH states. In the usual Hawking view, the situation is very different. The evaporation is expected to stop somehow slowly (when compared to the previous stages). Because the energy available inevitably becomes, at some point, smaller that it should be (in the sense that $M$ becomes smaller that the associated temperature $1/(8\pi M)$), the process slows down and the energy of the emitted particles decreases. In \cite{Barrau:2011md}, it was shown that this might be used as another discrimination tool between models.\\

It could also be that a periodicity with broader peaks does appear in the emitted spectrum, due to the ``large scale" structure of the area spectrum. This has been discussed in \cite{Agullo:2010zz}. In that case, the Hawking/LQG spectra could also be dicriminated for higher mass black holes \cite{Barrau:2011md}.  This possibility is however extremely unlikely, and we will not discuss it further, as a damping in the pseudo-periodicity is expected to take place \cite{G.:2008mj,FernandoBarbero:2011kb,Cao:2011zi}. \\

This analysis was pushed further in \cite{Barrau:2015ana} where recent results are accounted for. The fundamental excitations are now better understood as living on the horizon and as being elements of the Hilbert space of a SU(2) Chern-Simons theory \cite{Engle:2010kt,Engle:2010kt,Engle:2011vf}. The quantization of such a Chern-Simons theory with a compact gauge group is well-defined and the kinematical characteristics of a quantum black hole becomes quite clear \cite{Buffenoir:2002tx,Noui:2006ku,Noui:2006kv}. The role of the Barbero-Immirzi parameter $\gamma$ was studied in details and recovering the Bekenstein-Hawking entropy has been considered as a way to fix its value. It is however the coupling constant with a topological term in the action of gravity, with no consequence on the classical equations of motion. The strong dependence of the entropy calculation on $\gamma$ therefore remains controversial. Many progresses were recently made  \cite{Ghosh:2011fc,Ghosh:2013iwa,Solodukhin:2011gn,Geiller:2014eza,Achour:2014eqa}. The canonical ensemble formulation of the entropy making use of a quasi-local description shed a new light of the subject. The semi-classical thermodynamical properties can actually be recovered for any value of $\gamma$ if one assumes a non trivial  chemical potential conjugate to the number of horizon punctures.
A possible fundamental explanation to the exponential degeneracy would be to consider the area degeneracy as an analytic function of  $\gamma$ and to make an analytical continuation from real $\gamma$ to complex $\gamma$. This suggests that the quantum gravitational theory, defined in terms of self dual variables, could  account for the holographic degeneracy of the area spectrum of the BH horizon.\\

Two models of black holes were studied by a full MC simulation in \cite{Barrau:2015ana}. The first is based on the naive microcanonical view. It takes into account only the quantum geometry excitations, leading to  \cite{Agullo:2009eq}:
\begin{equation} 
\label{lqgbh}
S=\frac{\gamma_0}{\gamma} \frac{A}{4} + {o}(log(A)),
\end{equation} 
where $\gamma_0$ is of order one. Then, holographic black holes, where one uses the matter degeneracy suggested by quantum field theory with a cut-off at the vicinity of the horizon (that is an exponential growth of vacuum entanglement in terms of the BH area), were considered. The entropy becomes:
\begin{equation} 
\label{lqgbh2}
S=\frac{A}{4}+ \sqrt{\frac{\pi A}{6\gamma}} + o(\sqrt{A}).
\end{equation} 

\begin{figure}[!h]
\begin{center}
\includegraphics[scale=0.45]{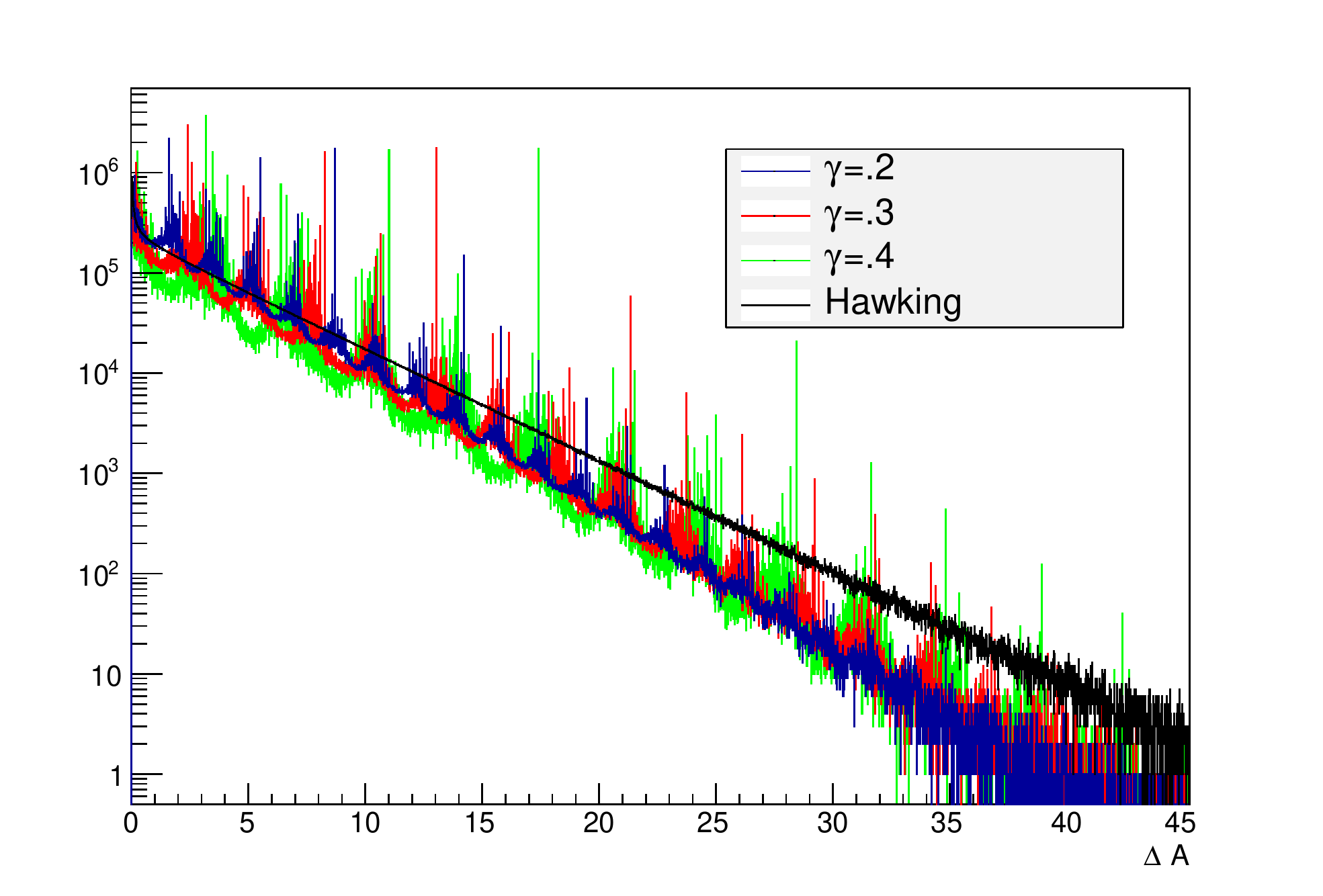}
\caption {Spectrum of a holographic black hole for different values of $\gamma$ as a function of $\Delta A$, from \cite{Barrau:2015ana}.} 
\label{fig1}
\end{center}
\end{figure}

\begin{figure}[!h]
\begin{center}
\includegraphics[scale=0.45]{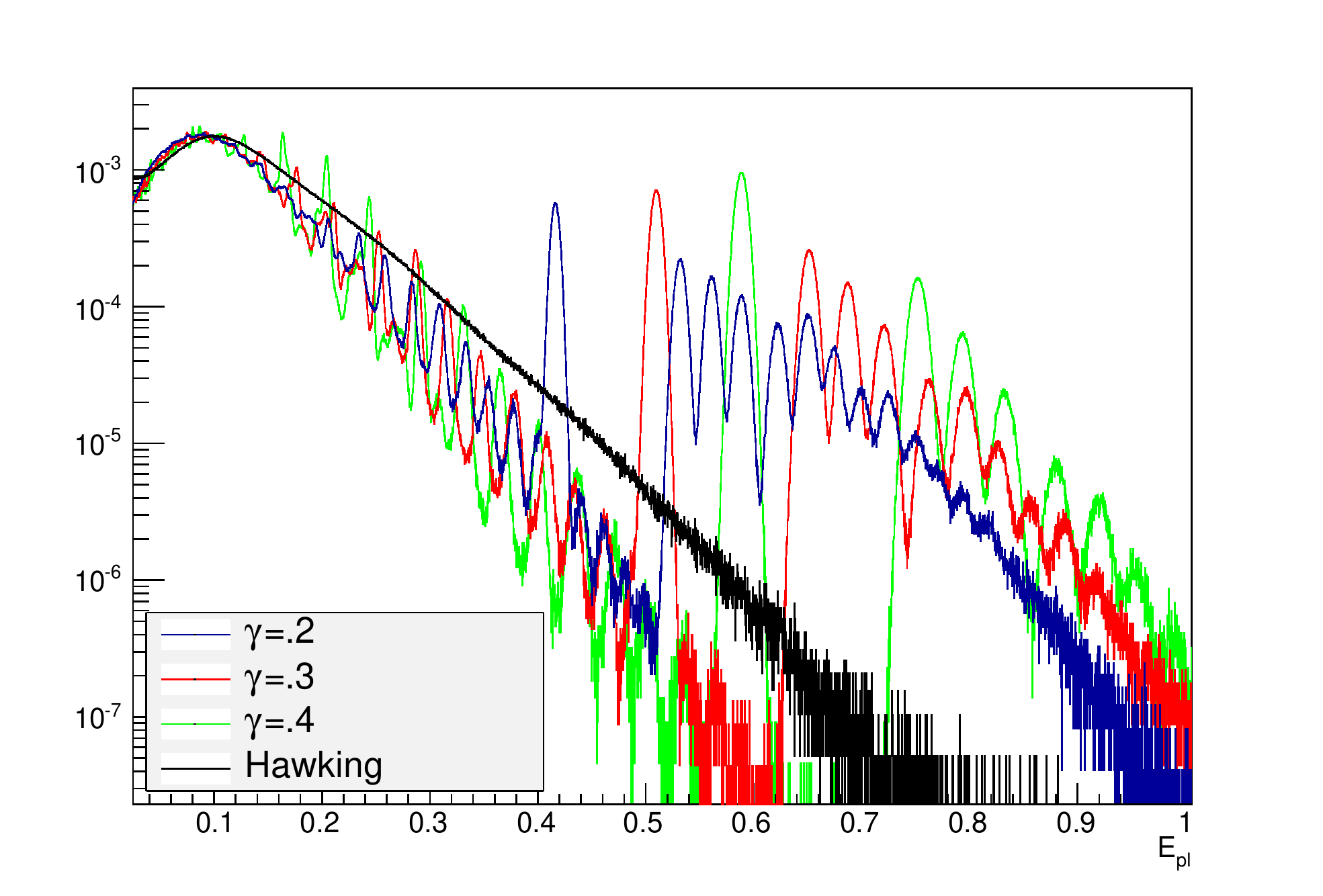}
\caption {$\gamma$ dependence of the integrated spectrum, as function of the energy of the emitted particle, in the holographic model, with a detector energy resolution of 5\%, from \cite{Barrau:2015ana}.} 
\label{fig2}
\end{center}
\end{figure}

The simulation has been performed with $10^7$ evaporating black holes. Fig. \ref{fig1} shows the results for different values of the Barbero-Immirzi parameter $\gamma$. This  $\gamma$ dependence interestingly shows up even though the leading order term of the black hole entropy, which mainly governs the transitions during the evaporation process, is {\it not} depending on  $\gamma$. This phenomenon is fully quantum gravitational in nature and is both due to the fact that  $\gamma$ enters in the discretization of the area spectrum and shows off in the sub-leading corrections to the entropy. The effects of a detector finite energy resolution are shown on Fig. \ref{fig2}.\\

This shows that the Hawking spectrum of a LQG BH has  two distinct parts: a nearly continuous background corresponding to the semi-classical stages of the evaporation and a series of discrete peaks associated with the deep quantum structure. Interestingly, $\gamma$ has an effect on both parts and becomes somehow measurable. In all cases, there are significant differences with the usual Hawking picture in the last stages.

\subsection{Greybody factors}

When dealing with evaporating black holes, a key element is the greybody factor -- closely related with the absorption cross section. The Hawking effect is approximated by a blackbody spectrum at temperature $T_H=1/(8 \pi M)$ with $M$ the mass of the BH. But the emitted particles have to cross a (gravitational and centrifugal) potential barrier before escaping to infinity. This induces a slight modification of the spectrum, captured by the cross section $\sigma$. The spectrum reads as:
\be
\frac{dN}{dt}= \frac{1}{e^{\frac{\omega}{T_H}} \pm 1} \sigma(M,s,\omega) \frac{d^3k}{(2 \pi )^3}, 
\ee

\noindent with  $s$ the particle spin, $\omega$ its energy, and $k$ its momentum. The cross section is, in general, given by
\be 
\sigma(\omega)_s= \sum_{l=0}^{ \infty} \frac{(2j+1) \pi}{\omega ^2} |A_{l,s} |^2,
\label{sigma}
\ee
where $A_{l,s} $ is the transmission coefficient of the mode with angular momentum $l$, and $j=l+s$ is the total angular momentum. It has been shown, in many different frameworks, to encode a lot on information on the chosen gravitational theory or on the underlying background spacetime. In the framework of LQG those cross sections have been studied only in \cite{Moulin:2018uap}.\\ 

The emphasis was put on BHs as described in  \cite{Alesci:2011wn,Modesto:2008im} where, instead of all {\it a priori} possible closed graphs, a regular lattice with edges of lengths $\delta_b$  and $\delta_c$ was chosen. The resulting dynamical solution inside the horizon is analytically continued to the region outside the horizon. Requiring that the minimum area is the one found in the LQG area operator spectrum, the model is reduced to one free parameter $\delta$, the so-called dimensionless polymeric parameter.  The effective LQG-corrected Schwarzschild metric is then given by:

\begin{eqnarray}
&& ds^2 = - G(r) dt^2 + \frac{dr^2}{F(r)} + H(r) d\Omega^2~, \nonumber \\
&& G(r) = \frac{(r-r_+)(r-r_-)(r+ r_{*})^2}{r^4 +a_o^2}~ , \nonumber \\
&& F(r) = \frac{(r-r_+)(r-r_-) r^4}{(r+ r_{*})^2 (r^4 +a_o^2)} ~, \nonumber \\
&& H(r) = r^2 + \frac{a_o^2}{r^2}~,
\label{g}
\end{eqnarray}
where $d \Omega^2 = d \theta^2 + \sin^2 \theta d \phi^2$, $r_+ = 2m$ and $r_-= 2 m P^2$ are the two horizons, and $r_* = \sqrt{r_+ r_-} = 2mP$, $P$ being the polymeric function defined by $P = (\sqrt{1+\epsilon^2} -1)/(\sqrt{1+\epsilon^2} +1)$, with $\epsilon=\gamma\delta$, and the area parameter $a_o$ is given by $a_0=A_{min}/8 \pi$. The parameter $m$ in the solution is related to the ADM mass $M$ by $M = m (1+P)^2$.\\

The case of massless scalar fields is quite easy to deal with. Since the BH is static and spherical, the field can be written as
$\Phi(r, \theta, \phi ,t)=R(r)S(\theta)e^{i(\omega t+m \phi)}$ and the generalized Klein-Gordon equation is

\be
\frac{1}{\sqrt{-g}}\partial_{\mu} (g^{\mu \nu }\sqrt{-g}\partial_{\nu} \Phi)=0,
\label{kg}
\ee

\noindent leading for the metric given in Eq. (\ref{g}), to the radial equation:

\be
\frac{\sqrt{GF}}{H} \frac{\partial }{\partial  r} \left( H\sqrt{GF}\frac{\partial  R(r)}{\partial  r}\right)+\left(\omega ^2 - \frac{G}{H} l(l+1) \right) R(r)=0.
\label{radialr}
\ee

\begin{figure}
\centering
    \includegraphics[width=.9\linewidth]{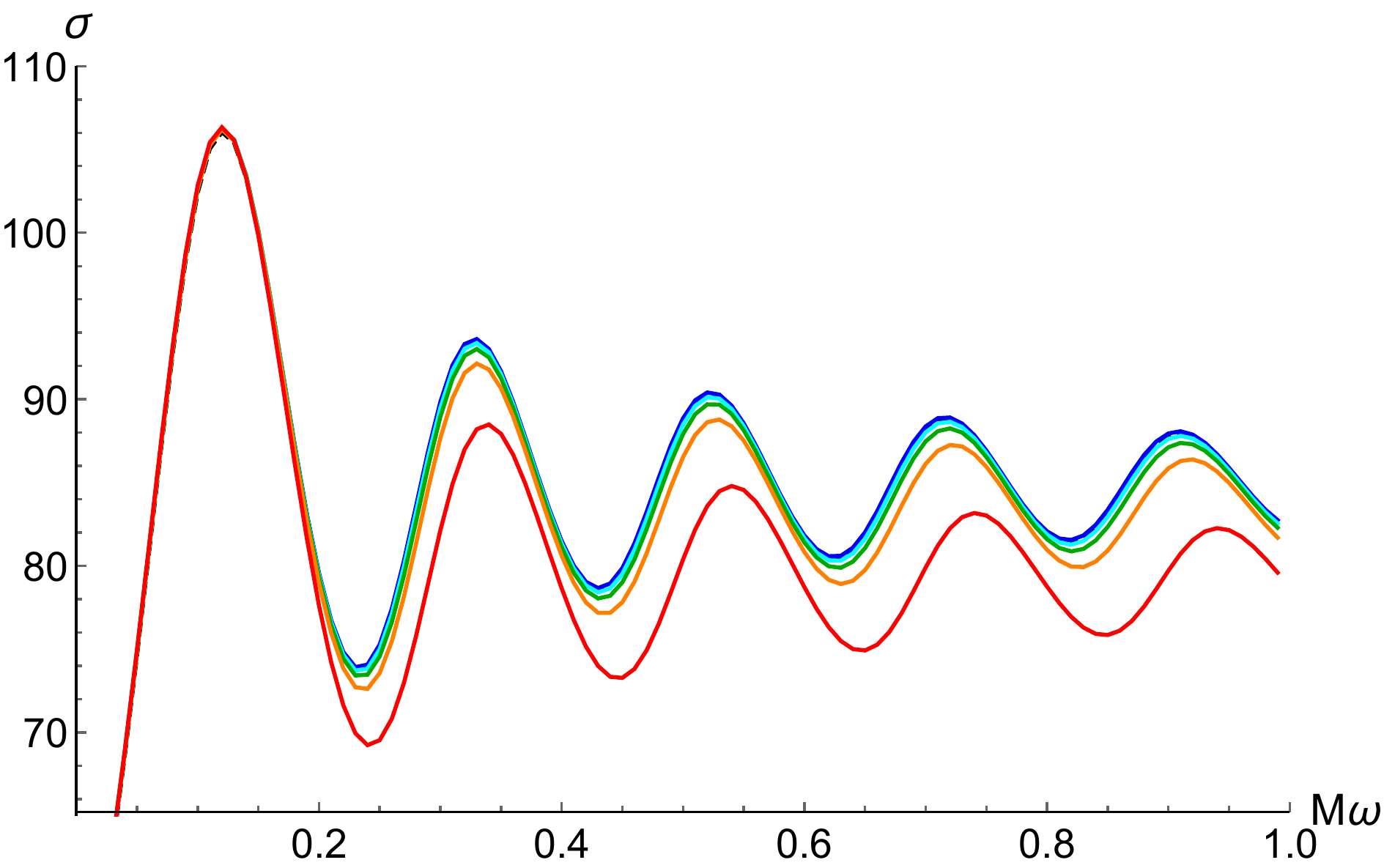}  
 \caption{Emission cross section for a scalar field with energy $\omega$ for a loop BH of mass $M$ for different values of $\epsilon$. From bottom to top: $\epsilon=10^{ \{-0.3,-0.6,-0.8,-1,-3\}}$. The blue line, corresponding to $\epsilon=10^{-3}$ is superposed with the cross section for a Schwarzschild BH. From \cite{Moulin:2018uap}.}
      \label{CSection}

\end{figure}

Using the tortoise coordinate $dr^{* 2} \equiv \frac{dr^2}{GF}$, one can impose the appropriate boundary conditions, fit the asymptotic solutions and sum over the different values of $l$ to get the final cross section, which is given in Fig. \ref{CSection}.

The cross section decreases when $\epsilon$ increases. One can also notice a shift of the pseudo-periodic oscillations toward a lower frequency (in $M\omega$). When $\epsilon < 10^{-0.8}$, it is hard to distinguish between the solutions. From the phenomenological viewpoint, it seems that taking into account the quantum corrections does not influence substantially the cross section of a scalar field for reasonable values of $\epsilon$ (that is $\epsilon \ll 1$). The main trend is however clear and if the actual value of $\epsilon$ happened to be unexpectedly high, it could be probed by a reduced cross section. \\

The case of fermions is more complicated and a specific derivation of the Dirac equation in the Newman-Penrose formalism had to be developed in \cite{Moulin:2018uap}. Basically, one is led to the following equation for the $R_+$ component of the Dirac spinor (the equation for $R_-$ is the conjugate):
\be 
\sqrt{HF}\mathcal{D} \left( \frac{\sqrt{HF}\mathcal{D}^{\dag} }{\lambda - i m_e \sqrt{H}} R_+ \right)  - (\lambda + i m_e \sqrt{H}) R_+ = 0,
\label{radialeq}
\ee

\noindent with $\mathcal{D}$ a radial operator

\be 
\mathcal{D}= \partial _r +  \left( \frac{G'}{8G} - \frac{F'}{8F}  \right) + \frac{iw}{\sqrt{GF}}.
  \ee
  
\begin{figure}
\centering
    \includegraphics[width=.9\linewidth]{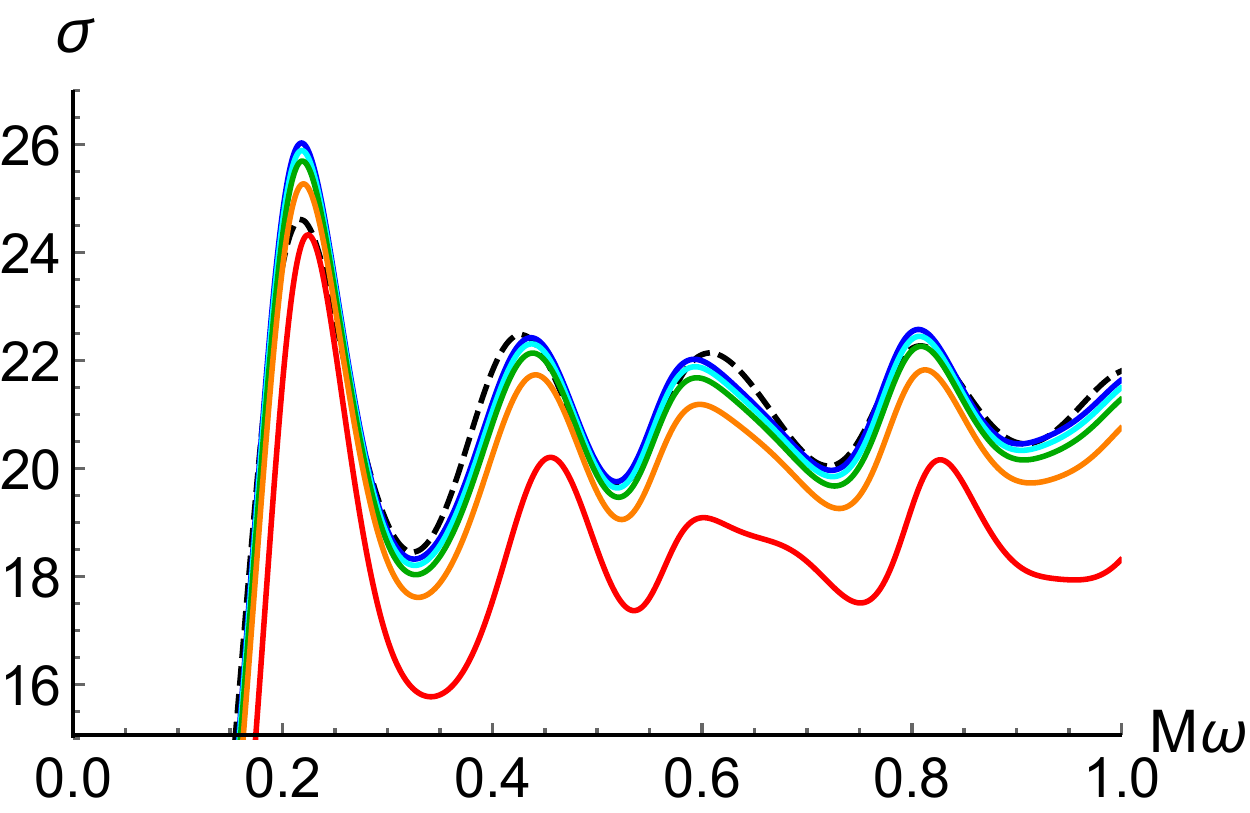}  
 \caption{Emission cross section for a fermionic field, with energy $\omega$, for a loop BH of mass $M$. From bottom to top: $\epsilon=10^{ \{-0.3,-0.6,-0.8,-1,-3\}}$. The dashed dark curve corresponds to the Schwarzschild cross section. From \cite{Moulin:2018uap}}
 \label{res}
 
 \end{figure}

\noindent The separation constant $\lambda$ is obtained by solving the angular equation, leading to $ \lambda^2=(l+1)^2$ for fermions. Results are given in Fig. \ref{res}. Once again, the general trend is a decrease of the cross section when the ``quantumness" increases. In addition, it was shown that the existence of a non-vanishing $a_0$ is the reason for the slight increase of the cross section on the first peak. The polymerization parameter and the minimal area do have different consequences.\\

 The considered polymerized model \cite{Alesci:2011wn} is just a first attempt and by no means a final statement on the quantum corrected geometry around an LQG BH. The same work on greybody factors should be carried out for models like \cite{BenAchour:2018khr,Ashtekar:2018lag,Ashtekar:2018cay}, to cite only a few. This however shows that some non-trivial features can be expected.

\subsection{Local perspective}

The previous view is based on the idea that the Hawking evaporation should be considered as a global phenomenon. The BH emits a particle and undergoes a transition from one area eigenstate to another one. When the BH is large, the density of states grows exponentially and reads (we make the Newton constant dependance explicit here) as $\rho(M)\sim \rm{exp}(M\sqrt{4\pi G/3})$, which means that the spectral lines are virtually dense in frequency for high enough masses. No quantum gravity effects are therefore expected well above the Planck mass. \\

This view is however not that straightforward. When the BH undergoes a transition from the mass $M_1$ to the mass $M_2$, which is extremely close to $M_1$ if the black hole is massive, the quantum state after the jump is -- in the global perspective -- completely different from the initial one. The final state corresponds to values of the spins (labelling the $SU(2)$ representations of the edges puncturing the horizon -- or colors of the graph) that are generically deeply different from the ones of the initial state. Assuming that the quasidense distribution of states is correct requires a full reassigning of the quantum numbers for every single transition, which is in tension with a quantum gravitational origin of the evaporation process. As we will explain later, if, instead, one assumes that the evaporation is due to a change of state of an ``elementary area cell", there is no reason for all of the other surfaces paving the horizon to change simultaneously their quantum state (as argued, {\it e.g.}, in \cite{Makela:2011vd}). This even raises a causality issue: how can a ``far away" elementary cell know how it should change to adjust to the others? \\

Another view, to account for this issue, was however suggested in \cite{Barrau:2016qri} (somehow in the line of \cite{Yoon:2012cq}), assuming that each particle emitted is basically due to the relaxation of the BH following a change of state of a single elementary cell. This was called a local quantum gravity dynamics. This does {\it not} assume that local process magically know the global BH quantities like temperature, entropy and mass: after the quantum jump, without any {\it a priori} knowledge of the picture, the BH relaxes through a semiclassical process consistent with the energy available. This naturally leads to a spectrum whose properties fit the Hawking description. 

This hypothesis leads to phenomenological results comparable to those of \cite{Bekenstein:1995ju} but with a clear foundation in the LQG framework. The key-point is that the same change of area $dA$ ($\sim A_{Pl}$) implies a relative peak separation in the spectrum $dE/T$ which is {\it independent} of the BH mass. Quantum gravity effects can therefore be expected to be measured for masses arbitrary far above the Planck mass. This deeply contrasts with what was believed to be expected in initial LQG studies. The density of {\it reachable} states is not quasi-dense anymore.\\

The eigenvalues of the area operator given by Eq. (\ref{eq1}) are not equally spaced: only in the large-$j$ limit does a regular line spectrum arise. It is shown in \cite{Barrau:2016qri} that this interesting feature could allow to distinguish between different LQG models of black holes (in particular those in the line of \cite{Ashtekar:1997yu} favoring low spin values and the holographic ones \cite{Ghosh:2013iwa} where higher spins could dominate). 

If one calls $nA_0/2$ the area variation associated with one quantum jump, $n$ being an integer and $A_0$ the basic area $\sim A_{Pl}$, the relative variation of energy of the emitted particles between emissions is
$
\Delta E/E\approx nA_0/(2A).
$
The change in energy is therefore negligible and the line structure should be observable if it exists: the BH mass evolution  during its evaporation does not erase out this feature.

The criterion for the detection of a signal coming from an evaporating primordial black hole (PBH) \cite{Carr:2009jm} consists in asking for a mean time $\Delta t$ between two measured photons smaller than a given reference time interval $\Delta t_0$. This allows to estimate a maximum distance for detection of
\begin{equation}
R_{max}\approx\sqrt{\frac{S\Delta t_0}{M}}.
\end{equation}

The realistic case however corresponds to the signal emitted by a distribution of PBHs with different masses. Does the global line structure remains? It was shown that if the temperature of the universe does not change by more that 5-10 \% during the formation of the considered PBHs, the line structure holds.\\ 

Another issue had to be considered seriously: when the temperature of the BH is higher than the quantum chromodynamics (QCD) confinement scale, the evaporating BH also emits partons that will fragmentate into hadrons. Some of those will then decay into gamma-rays, denoted as ``secondary". The secondary instantaneous spectrum reads as

\begin{eqnarray}
      \frac{d^2N_{\gamma}}{dEdt}&=&
      \sum_j\int_{Q=E}^{\infty}\alpha_j\Gamma_j(Q,T)
      \left(e^{\frac{Q}{T}}-(-1)^{2s_j}\right)^{-1}\\ \nonumber
      &\times&\frac{dg_{j\gamma}(Q,E)}{dE}{dQ},
\end{eqnarray}
where $j=1,...6$ is the flavor, $s_j=1/2$ $~ \forall j$, $dg(Q,E)/dE$ is the normalized differential fragmentation function (determined using the ``Lund Monte Carlo" PYTHIA code \cite{Sjostrand:2014zea}), $Q$ being the quark energy, $T$ the temperature of the black holes, $\alpha$ the number of degrees of freedom, $\Gamma$ the cross section, and $E$ the photon energy. The time-integrated spectrum is then given by
\begin{equation}
\frac{dN_{\gamma}}{dE}=\int_{M_i}^{M_f}\frac{d^2N_{\gamma}}{dEdt}\frac{dt}{dM}dM.
\end{equation}

Those secondary photons will obviously not exhibit the line structure of quantum gravitational origin. The numerical simulation performed in 
\cite{Barrau:2016qri} however shows that, quite surprisingly, those  electromagnetic quanta are not numerous enough  to wash-out the primary signal and its line structure, which could indeed still be measured.\\\

If this local view for the evaporation of black holes is correct, this means that this should lead to a line structure in the spectrum, even arbitrarily far away from the Planck mass.

\section{Bouncing black holes}

\subsection{The model}

Recently, the possibility that black holes could actually be bouncing objects has been revived.  In its current ``LQG-compatible" version, the model was first introduced in \cite{Rovelli:2014cta}, and its consequences were studied in \cite{Barrau:2014hda}. It was then refined in \cite{Haggard:2014rza,Haggard:2015iya}. Basically, the idea is that what happens to the Universe in LQC, that is a bounce, should also happen to black holes. As the contracting Friedmann solution is connected to the expanding one by a quantum tunneling, the classical black hole solution is expected to be glued to the white hole one by quantum gravitational effects. It is in line with other works based on different assumptions, {\it e.g.} \cite{Giddings:1992hh,Hajicek:2001yd}. The process takes a time proportional to $M^2$, whereas the Hawking process requires a time of order $M^3$. Black holes would therefore bounce before they do evaporate and the Hawking radiation would be seen as a kind of a dissipative correction. \\

\begin{figure}
\centering
    \includegraphics[width=.9\linewidth]{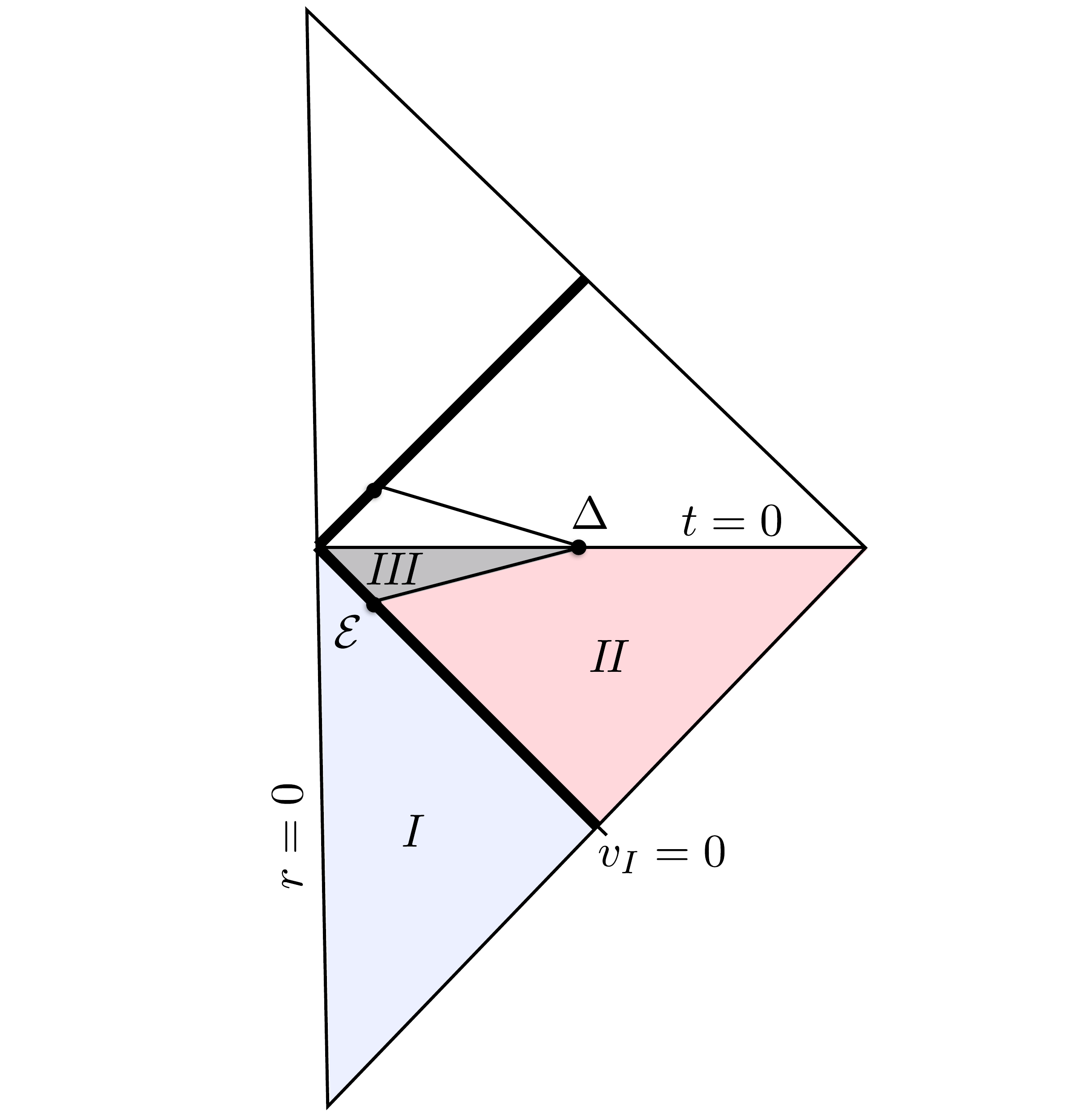}  
 \caption{Causal diagram for a bouncing black holes, from \cite{Haggard:2014rza}. (I) is flat, (II) is Schwarzschild, ans (III) is the ``quantum gravity" region.}
 \label{penrose}
 \end{figure}

The important result of \cite{Haggard:2014rza} is that a metric exists for a bouncing black-to-white hole. It is a solution to the Einstein equations outside a finite region and beyond a finite time duration.  This means that it is possible to have a bounce from a black hole into a white hole without any spacetime modification at large radius. The quantum region extends slightly outside the Schwarzschild radius and can have a short duration. The associated Penrose diagram is shown in Fig. \ref{penrose}.

Because of the gravitational redshift the bounce is seen as nearly ``frozen" by a distant observer but it is extremely fast for a clock comobile with the collapsing null shell. In this sense, a BH is a star that is collapsing and bouncing seen at very slow motion from the exterior. \\

The key-point  is to assume that classicality might not be determined by, {\it e.g.}, the Kretschmann invariant (${\cal R}^2=R^{abcd} R_{abcd}$) but by 
\be
           q  = l^{2-{b}}_P  \ {\cal R} \ \tau^{b}, 
\ee
with $b$ of order unity and $\tau$ is the (asymptotic) proper time. In this expression units have been reinserted for clarity. This opens the door to a possible cumulative effect like in the decay of an unstable nucleus.\\

The metric is entirely determined by two functions of $u$ and $v$,
\be
ds^2=-F(u,v) du dv + r^2(u,v)(d\theta^2+\sin^2\theta d\phi^2),
\ee
whose explicit expression has been calculated in \cite{Haggard:2014rza}. Interestingly, this also means that strong quantum gravity effects may appear outside the event horizon (which becomes, in this context, a trapping horizon) at $R=(7/6)R_S$ \cite{Haggard:2016ibp}. As far as this study is concerned, the key-point is that the bouncing time is given by $\tau=4kM^2$ (although this expression is hard to recover from the full theory \cite{Christodoulou:2016vny}). The k parameter has a lower bound ($k>0.05$) and will be varied in the next sections. 

\subsection{Individual events and fast radio bursts}

The question of the detectability of those bouncing black holes naturally arises. At this stage, a detailed model for the emission from the white hole is missing. Two hypothesis can however reasonably by made. 

The first one is simply based on dimensional analysis. The hole size is the only scale of the problem. It is therefore expected that the wavelength of the emitted radiation is of the order of the bouncing BH diameter. This makes clear sense and this is in agreement with what happens, {\it e.g.}, during the Hawking evaporation. The associated signal is called the low-energy component.

The second hypothesis relies on the symmetry of the process (this might not be completely true \cite{DeLorenzo:2015gtx}, but this does not change the argument). What goes out of the white hole is what went in the black hole. In this model, the bouncing star is formed by a collapsing null shell. The energy of the emitted radiation should therefore be the same than the one of the incoming photons. If we consider PBHs formed in the early universe by the collapse of over-densities the correspondance between the mass and the time is known. And time is also in one-to-one correspondance with the temperature of the Universe. So, for a given BH mass, one can calculate the energy of the emitted radiation, called the high-energy component. \\

The idea of explaining fast radio bursts (FRBs) by bouncing black holes was suggested in \cite{Barrau:2014yka}. Basically, FRBs are intense radio signals with a very brief duration. Events were, among others, observed at the Parkes radio telescope \cite{Lorimer:2007qn,Keane:2012yh,Thornton:2013iua} and by the Arecibo Observatory \cite{Spitler:2014fla}. Could they be explained by (the low-energy component of) bouncing black holes ?

As mentioned before, the bouncing time can be estimated to be of the order of
\be
  \tau=4k\  M^2.
  \label{tau}
\ee

For the phenomenology of FRBs, one sets the parameter to its lowest possible value: $k=0.05$. PBHs with an initial mass around
\be
M_{t_H}= \sqrt{\frac{t_H}{4k}}\sim 10^{26}~{\rm g},
\label{mass}
\ee 

where $t_H$ is the Hubble time, would therefore be expected to explode today.  One can notice that, naturally, this mass is much higher that $M_{\star}\sim 10^{15}~{\rm g}$ corresponding to black holes that would require a Hubble time to evaporate by the Hawking process. In the case of the low energy channel of bouncing BH, the emitted radiation wavelength should be of the order of 200 microns, three orders of magnitude below the measured 20 cm of FRBs.\\

This apparent discrepancy has been addressed and solved in \cite{Barrau:2018kyv}. The key idea lies in the fact that if the black-to-white hole transition is to be understood as a tunneling process, the lifetime of a BH should be considered as a random variable. The probability for a black hole not to have bounced after a time $t$ is given by
\begin{equation}
P(t)=\frac{1}{\tau}e^{-\frac{t}{\tau}}.
\end{equation}
Let us model the shape of the signal emitted by a single black hole by a simple Gaussian function of width $\sigma_E$.
The full signal due to a local distribution of bouncing black holes is given by 
\begin{equation}
\frac{dN_{\gamma}}{dE}=\int_{M_{Pl}}^{\infty}Ae^{-\frac{(E-E_0)^2}{2\sigma_E^2}}\cdot \frac{dN}{dM}(M)\cdot\frac{1}{4kM^2}e^{-\frac{t_H}{4kM^2}}.
\end{equation}

The key-point is that the mean energy of the detected signal is {\it not} necessarily the naively expected one, that is may {\it not} be $E\sim1/(4M_{t_H})$ where $M_{t_H}$ is such that $t_H=4kM_{t_H}^2$ (this corresponds to BHs having a characteristic lifetime of the order of the age of the Universe, leading to the emitted wavelength 3 orders of magnitude too small to account for FRBs). If the mass spectrum of PBHS is however peaked around a mass $M_0$,
\begin{equation}
\frac{dN}{dM}\propto e^{-\frac{(M-M_0)^2}{2\sigma_{\footnotesize{M}}^2}},
\end{equation}
which can be different than $M_{t_H}$, the mean emitted energy will be around $1/(4M_0)$ which can differ from $1/(4M_{t_H})$. This happens because of the distributional nature of the bouncing time.\\

\begin{figure}
\centering
    \includegraphics[width=.9\linewidth]{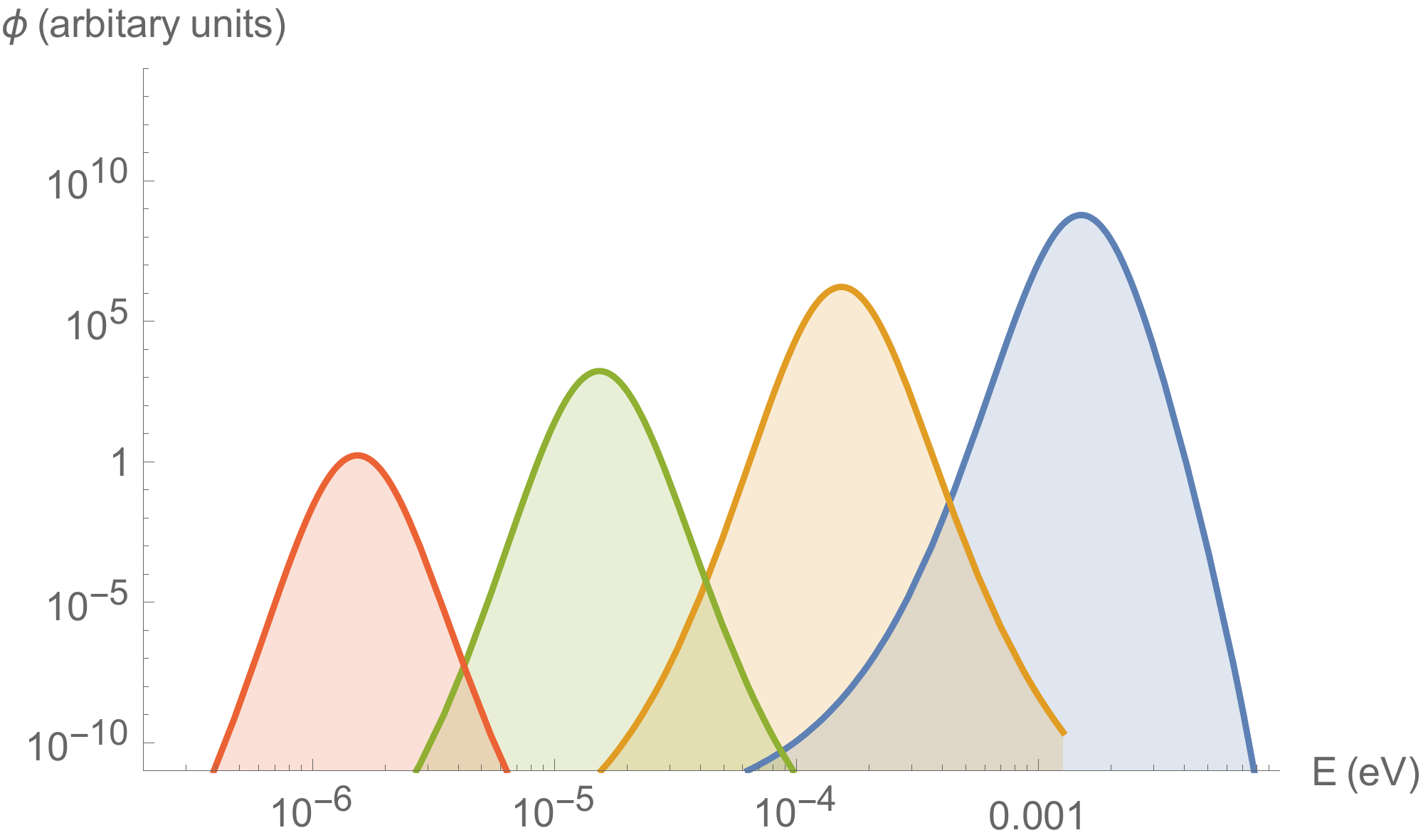}  
 \caption{Electromagnetic flux emitted by bouncing BHs for a mean mass $M_0$ of (from right to left) $M_{t_H}$, $10M_{t_H}$, $100M_{t_H}$, and $1000M_{t_H}$, normalized such that the total mass going into PBHs is the same. From \cite{Barrau:2018kyv}.}
 \label{sto}
 \end{figure}
 
 In Fig. \ref{sto}, the emitted photon flux is shown for different values of the mean mass $M_0$ of the mass spectrum: $M_{t_H}$, $10M_{t_H}$, $100M_{t_H}$, and $1000M_{t_H}$. This shows that the energy of the radiation does depend on this value, even if the parameters of the model are otherwise fixed. Since a given mean lifetime $\tau=4kM^2$ does not imply a fixed expected energy, the three orders of magnitudes needed to match the measured energy of FRBs can be accounted for with a mass $M_0 = 1000M_{t_H}$, which corresponds to the left curve in Fig. \ref{sto}. \\

\begin{figure}
\centering
    \includegraphics[width=.9\linewidth]{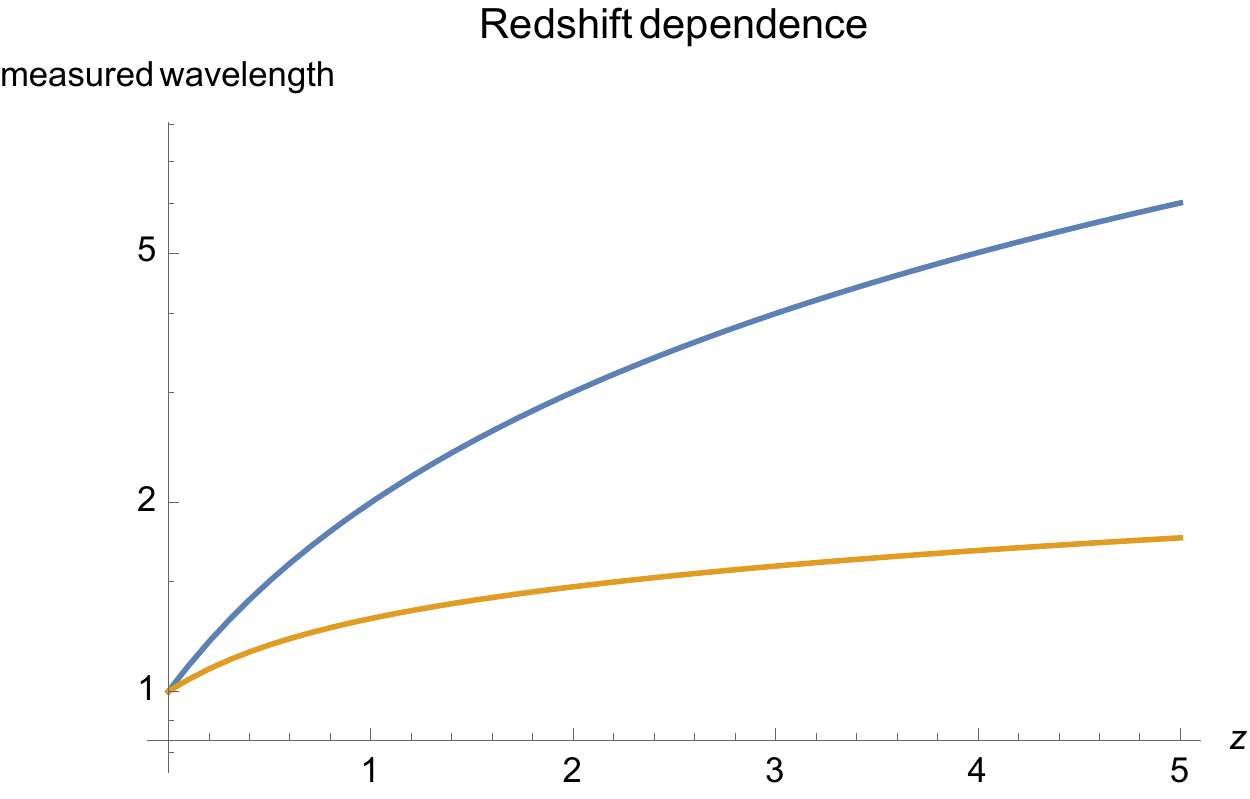}  
 \caption{Measured wavelength, normalized to the rest-frame one, as a function of the redshift. The upper curve is for a conventional astrophysical signal and the lower one is for bouncing black holes. Reproduced from \cite{Barrau:2016fcg}, with the permission of AIP Publishing.}
 \label{red}
 \end{figure}

This explanation for FRBs is unquestionably exotic when compared to more conventional astrophysical interpretations (especially when considering that one ``repeater" has been observed -- it could however well be that there are different populations of FRBs). What makes the scenario however meaningful is that it is testable, due to a specific redshift dependance. When observing a galaxy at redshift $z$, the measured energy of the signal emitted by any astrophysical object (including decaying dark matter) will be $E/(1+z)$ for a rest-frame energy $E$. This is not the case for bouncing black holes: BHs that have bounced far away and are observed today had a shorter bouncing time and consequently a smaller mass. The energy of the emitted radiation is therefore higher and this compensates for the redshift effect. The observed wavelength of the signal from an object at redshift $z$ can be written as:
\begin{eqnarray}
\lambda_{obs}^{BH}\!&\sim&\! \frac{2Gm}{c^2} (1+z)\ \ \times \\ \nonumber
&& \  \sqrt{\frac{H_0^{-1}}{6\,k\Omega_\Lambda^{\,1/2}
}\ \sinh^{-1}\!\!\left[ \left(\frac{\Omega_\Lambda}{\Omega_M}\right)^{\!1/2} (z + 1)^{-3/2}\right]},
\label{redhisft}
\end{eqnarray}
where we have reinserted the physical constants; $H_0, \Omega_\Lambda$ and $\Omega_M$ being respectively the Hubble rate, the cosmological constant, and the matter density. This is to be contrasted with what happens for standard sources whose measured wavelength is related to the observed wavelength by
\begin{equation}
\lambda_{obs}^{other}=(1+z)\lambda_{emitted}^{other}~,
\end{equation}
as shown in Fig. \ref{red}.\\

Importantly, is was also shown in \cite{Barrau:2018kyv} that even if the mass spectrum is wide, it could still be possible to explain FRBs. It could be that most bouncing BHs lead to a signal of wavelength 0.02 cm and that only the tail (whose existence is due to the probabilistic nature of the lifetime) of the distribution is actually detected by radio-telescopes. If the real emission peak is in the infrared band – which should naturally occur if the mass spectrum is, itself, not peaked  – it could very well be that it is just unobserved today. Observatories in the infrared  have time constants that are too high to allow for the measurement of fast transient phenomena and no large survey is being carried out. In this case, a prediction of the model is that one should expect a higher flux as the energy increases.\\

Finally, it is worth considering the high-energy emission. The bouncing BHs then act as ``redshift freezing machines" for collapsing fields which are emitted back at the energy they had when being absorbed. However, in the meanwhile, the age of the surrounding universe has grown tremendously. In simple models, PBHs form with a mass of the order of the Hubble mass at the formation time. For BH masses as considered here (around $10^{26}$ g), this corresponds to a temperature of the Universe around the TeV. New very high energy telescopes, like the Cherenkov Telescope Array (CTA) could detect bursts is this energy range, as suggested by this model. \\

The redshift dependance for this component is qualitatively the same than for the low-energy one, but for different reasons. For a BH exploding at redshift $z$ and cosmic time $t$, the energy is determined by the temperature of the universe when the formation took place. It is proportional to the inverse square root of the time which is in turn proportional to the horizon mass, that is to the BH mass. So, the emitted wavelength is proportional to the square root of the mass of the BH. This leads to an observed wavelength 
\begin{equation}
\lambda_{obs}\propto(1+z)\left(\sinh^{-1}\!\!\left[ \left(\frac{\Omega_\Lambda}{\Omega_M}\right)^{\frac12}\! (z + 1)^{-\frac32} \right]\right)^{\frac{1}{4}}.
\end{equation}
As previously stated, this is a flatter dependance than for astrophysical effects. \\

\begin{figure}
\centering
    \includegraphics[width=.9\linewidth]{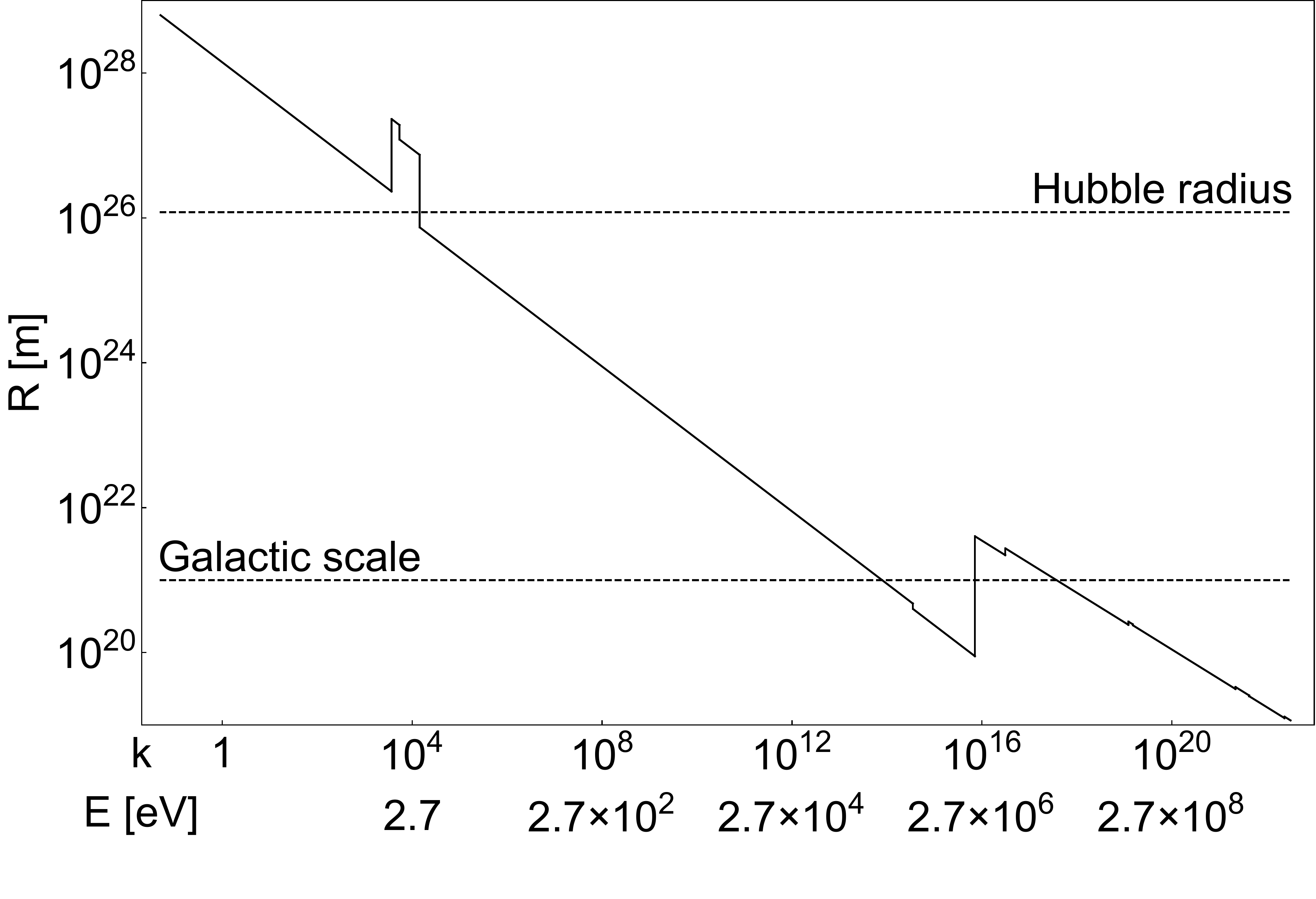}  
 \caption{Maximum distance at which a single bouncing BH can be observed through its low-energy component, as a function of the $k$ parameter, from \cite{Barrau:2015uca} (Copyright IOP Publishing.  Reproduced with permission.  All rights reserved).}
 \label{dist}
 \end{figure}

It is meaningful to evaluate the maximal distance at which one could observe a bouncing black hole. This question was addressed in \cite{Barrau:2015uca}, allowing the $k$ parameter, which determines the bouncing time, to vary. The minimum value of $k$ is such that the quantum effects have enough time to make the bounce happen and the maximum is such that the bouncing time remains smaller than the Hawking time. The study was carried out taking into account the size of the detector (and its detection efficiency), the absorption during the propagation over cosmological distances, and the number of measured photons required for the detection to be statistically significant. As $k$ increases, the global trend is a decrease of the maximum distance at which the bouncing BH can be observed. This comes both from the fact that BHs are lighter for higher values of $k$ (for a given bouncing time) and from the fact that they emit higher energy (and therefore fewer) particles. However, quite subtle effects also appear. For example, the distance can slightly decrease above the threshold of emission of a new stable particle (leaving less energy available for the considered photons) whereas it can increase when new particles decaying into gamma-rays are produced. For $k$ varying between 0.05 to $10^{22}$, the maximum detectable distance varies from the Hubble scale to $10^{19}$ m for the low-energy component, as shown in Fig. \ref{dist}, and from $10^{24}$ to $10^{16}$ m for the high energy component. 

\subsection{Background}

It is also important to consider a possible background emission. In this case, one does not look for a single event but from the diffuse emission due to a distribution of BHs. The number of photons detected per time unit, surface unit, and energy unit is given by:
\begin{equation}
\frac{dN_{mes}}{dEdtdS}=\int\Phi_{ind}((1+z)E,R) \cdot n(R) \cdot A(E) \cdot f(E,R)  dR,
\label{flux_int}
\end{equation}
where $\Phi_{ind}(E,R)$ is the flux emitted by a single BH at distance $R$ and at energy $E$, $n(R)$ is the number of BHs bouncing at distance $R$ per unit time and volume, $A(E)$ is the acceptance of the detector convoluted with its efficiency and $f(E,R)$ is the absorption. The $n(R)$ term does depend on the shape of the initial mass spectrum of PBHs which is unknown. It has however been checked that varying this shape has no significant impact on the results.\\

The study was carried out for both the low-energy and the high-energy components. In this latter case it is important to take into account the hadronization of emitted quarks that will produce hadrons potentially decaying into gamma-rays. This was modeled using the PYTHIA Monte-Carlo program \cite{Sjostrand:2014zea}. Quite surprisingly, the result is that, due to a kind of redshift-compensation effect, the integrated signal is very similar to the single event one. It basically appears as a distorted gaussian function  \cite{Barrau:2015uca}.\\

This also raised the question to whether it could be possible to explain the gamma-ray excess coming from the galactic center, as observed by the Fermi satellite. This has been reported in \cite{Hooper:2010mq,Abazajian:2012pn,Gordon:2013vta} and even observed at higher galactic latitudes \cite{Gordon:2013vta,Daylan:2014rsa}. Once again many astrophysical interpretation have been suggested. Millisecond pulsars are probably the most convincing hypothesis (see, {\it e.g.}, \cite{Bartels:2015aea}), it is however not yet fully satisfactory \cite{Daylan:2014rsa} and their is room for new physics. Interestingly, it was demonstrated in \cite{Barrau:2016fcg} that bouncing BHs can indeed explain the Fermi excess if the $k$ parameter is chosen at its higher possible value. It is worth noticing that the values required to explain either the FRBs or the GeV gamma-ray excess are not ``random" but either the smallest or the highest possible ones. \\

In \cite{Barrau:2016fcg}, the secondary spectrum, mostly due to the decay of neutral pions, was shown to be well approximated by 
\begin{equation}
f(E,\epsilon)=\frac{a\epsilon^b}{\pi\gamma}\left[\frac{\gamma^2}{(\epsilon-\epsilon_0)^2+\gamma^2}\right] e^{-\left(\frac{4\epsilon}{E}\right)^3},
\end{equation}
$E$ being the quark energy, $\epsilon$ the photon energy, $a=50.7$, $b=0.847$, $\gamma=0.0876$ and $\epsilon_0=0.0418$ (the energies being in GeV), whereas the direct emission due to the low-energy component (the high-energy component cannot be smaller than a TeV and is not relevant for this study) is given by
\begin{equation}
g(E,\epsilon)=Ae^{-\frac{(\epsilon-E)^2}{2\sigma^2}}+3N\sqrt{2\pi}A\sigma f(E,\epsilon),
\end{equation}
where $N$ is the number of flavors of quarks with $m<E$.\\

\begin{figure}
\centering
    \includegraphics[width=.9\linewidth]{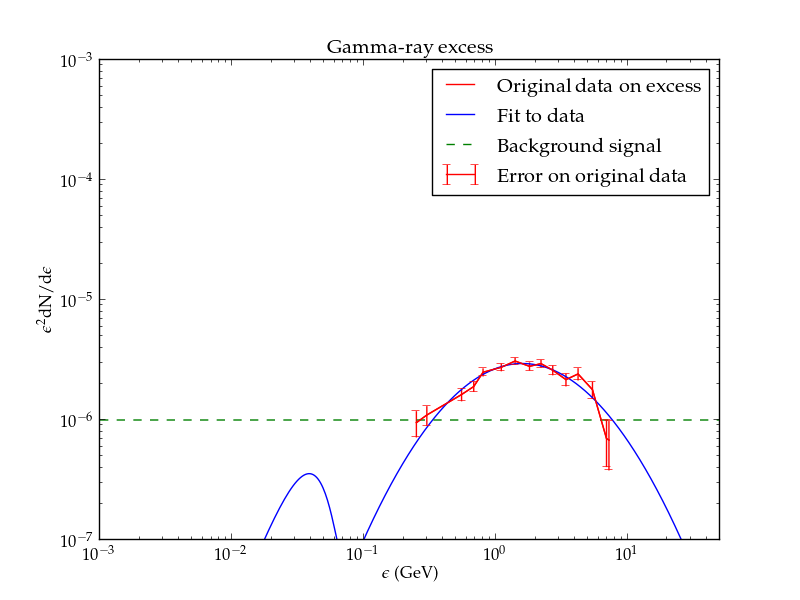}  
 \caption{Fit to the Fermi excess with bouncing black holes. Reprinted from \cite{Barrau:2016fcg}.}
 \label{fermi}
 \end{figure}

The best fit is shown in Fig. \ref{fermi}. The fact that the bouncing BH signal can account for the data is in itself non-trivial. It is, for exemple, absolutely impossible to reproduce the measurements with evaporating BHs. In addition, the most important result here lies in the amplitude of the little bump on the left of the plot. It is associated with the secondary emission (that is the one coming from the hadronization and subsequent decay of emitted partons). As the number of emitted quarks and gluons is much higher than the number of directly emitted photons (responsible for the main bump) it could have been (wrongly) expected that this indirect emission conflicts with the background displayed as the horizontal green dashed line on the plot. Due to the subtle energy distribution in the jets, this is not the case and, at this stage, the explanation by bouncing BHs does work satisfactorily. 

\section{Dark Matter}

The idea that if bouncing BHs are a substantial part of dark matter (DM), this might have an effect on galaxy clustering was introduced in \cite{Raccanelli:2017xee}. Several possible constraints were considered.\\

Only recently, however, was a new scenario for the evolution of black holes proposed \cite{Bianchi:2018mml}, with possible important consequences for DM. In this model, a black hole first evaporates, according to the usual Hawking process. When it becomes Planckian the tunneling probability to turn into a white hole, estimated to be of the order of
\begin{equation}
P\sim e^{-\frac{M^2}{M_p^2}},
\label{eq1}
\end{equation}
becomes large. Old black holes have a large interior volume \cite{Christodoulou:2014yia}: even if the Schwarzschild radius is fixed the ``physical" volume available inside does increase with time. This remains true for the formed white hole, although its mass is small (the volume is of the order of $M_i^4$ where $M_i$ is the initial mass). The white hole lifetime is also of the order of $M_i^4$. This scenario meets the conditions required to solve the information paradox.

It can be seen as a ``less radical"  proposal than the one presented in the previous sections. There is still a black to white hole transition, in agreement with the arguments given before, but instead of the very small bouncing time $M_i^2$, it takes the time suggested by the usual instanton solution. This is probably a more conservative and natural scenario.\\

In addition, following \cite{MacGibbon:1987my}, it was suggested that dark matter could be formed by such white hole relics \cite{Rovelli:2018hbk}. In \cite{Rovelli:2018okm}, the central argument is pushed forward. For those objects to
be still present in the contemporary Universe, one needs their lifetime to be larger than the Hubble time $t_H$, that is
\begin{equation}
M_i^4>t_H.
\label{eq2}
\end{equation}
On the other hand, for those relics to be formed by evaporated black holes, one needs
\begin{equation}
M_i^3<t_H,
\label{eq3}
\end{equation}
where $M_i^3$ if the Hawking evaporation time. This leads to
\begin{equation}
10^{10}~{\rm g}<M_i<10^{15}~{\rm g}.
\label{eq4}
\end{equation}
It is argued in \cite{Rovelli:2018okm} that this correspond to typical Hubble masses at reheating, making the scenario convincing. \\

It should be emphasized that quite a lot of models leading to stable relics at the end of the Hawking evaporation process have been proposed so far, relying on many different assumptions (see \cite{Barrow:1992hq,Zeldovich:1983cr,Aharonov:1987tp,Banks:1992ba,Banks:1992is,Bowick:1988xh,Coleman:1991jf,Lee:1991qs,Gibbons:1987ps,Torii:1993vm,Callan:1988hs,Myers:1988ze,Whitt:1988ax,Alexeyev:2002tg} to mention only a few historical references among many others). In those models, the relics are completely stable. This makes the situation easier: the only constraint is then that initial primordial black holes did evaporate within the Hubble time but there are no lower bound on $M_i$. From this point of view, the new model \cite{Rovelli:2018okm} is more challenging than the usual pictures, which does not make it wrong.\\

The key-point is of course to find a way to produce enough primordial black holes so that the white hole relics account for dark matter, without relying on too exotic physics as this is one of the motivation for this new scenario. As the CMB-measured amplitude and slope of the primordial power spectrum would lead to a vanishingly small number of primordial black holes, an extra input is obviously needed. A possibility would be to follow \cite{Barrau:2003xp} and use Starobinsky's broken scale invariance spectrum \cite{Starobinsky:1992ts}.  The main idea is that power is increased at small scales through a step in the power spectrum. 

Let us call $M_{H,e}$ the Hubble mass at the end of inflation, $p^2$ the ratio of the power on large scales with respect to that on small scales, $\delta_{min}$ the minimum density contrast required to form a black hole, $M_{WH}$ the mass of the white hole, and $\Omega_{WH,0}^2$ the abundance of white holes today, $LW$ the Lambert-W function, and $\sigma_{H}$ the mass variance. One can then show that
\begin{equation}
p\approx\frac{\sigma_{H}^{CMB}}{\delta_{min}}\sqrt{LW\left\{\frac{8.0\times
10^{-6}}{2\pi \Omega_{WH,0}^2}
\left[ \frac{M_{WH}}{M_p} \right]^2
\left[ \frac{10^{15}~{\rm g}}{M_{H,e}} \right]^3\right\}}.
\label{eq5}
\end{equation}
Requiring $\Omega_{WH,0}\approx 0.3$, $\delta_{min}\approx 0.7$ and $M_{WH}\approx M_p$ allows one to perform an explicit evaluation of $p$ and this fixes the parameters of the scenario assuming that the reheating temperature is high enough.\\

However, a major problem remains to be solved. If the white hole relics are to be made by primordial black holes with initial masses between $10^{10}~{\rm g}$ and $10^{15}~{\rm g}$, one must consider the sever constraints associated with nucleosynthesis. The $D/H$, $Li^6/Li^7$, and $He^3/D$ ratio mustn't be distorted by the evaporation of black holes (assumed to be the ``seeds" of the white hole relics) beyond observed values \cite{Carr:2009jm}. This forbids the easy formation of enough relics, unless a way to evade those constraints is found. This is the major challenge for future studies (which is fortunately easier to deal with when extended mass functions are taken into account \cite{Carr:2017jsz}).\\

Another possibility was imagined in \cite{Rovelli:2018hba}. Here, the objets are assumed to be formed before the bounce in a cosmological model where the Big Bang singularity is replaced by a tunneling between the classically contracting and the classically expanding Friedmann solutions, as suggested by loop quantum cosmology \cite{lqc9}. This is in principle consistant and other theories of quantum gravity might lead to this new paradigm. This evades the previously mentioned problem. Although in a different setting, the possibility was already considered in \cite{Carr:2011hv,Clifton:2017hvg,Carr:2017wkz}.

Very interestingly, the proposal is also related with the idea that the entropy and arrow of time could be perspectival \cite{Rovelli:2015dha}. This approach sheds a new light on the old paradox of the apparently low entropy of the initial state of the Universe: the Universe is not anymore homogeneous at the bounce and our observed entropy is determined by the fact that we cannot access the huge volume inside the abundant white hole remnants. It might seem puzzling that the authors explain the ``un-naturally" low entropy of the Universe by arguing that the probability for us to be where we are (outside of a relic) is only one part in $10^{120}$. It is however meaningful in the sense that a special position is much more anthropically ``acceptable" than a special state.\\

More importantly, it seems hard for the dark matter remnants to be already present at the bounce time. The current density of the universe is $\rho_0\sim 10^{-30}~{\rm g}~{\rm cm}^{-3}$. If we assume the usual cosmological evolution, we had at least 60 e-folds of inflation followed by approximately 60 e-folds of radiation, matter, and cosmological constant dominated expansion. It means that the scale factor has increased by at least a factor $10^{52}$ since the bounce. The density of remnants should then be at least $10^{156}\times10^{-30}=10^{126}~{\rm g}.{\rm cm}^{-3}$ at the bounce, that is $10^{33}\rho_{Pl}$. Leaving appart the fact that this value is probably unphysical (the bounce would have happened before when thinking in the positive time direction), this is anyway incompatible with Planck mass and Planck size remnants (that cannot lead to a density higher that the Planck density without merging).

This could be evaded by assuming that no inflation took place but this would require a quite exotic cosmological evolution. A nice feature of bouncing models is precisely to be compatible with inflation \cite{Ashtekar:2011rm,bl,Linsefors:2014tna,Bolliet:2017czc,Martineau:2017sti}. However, a possible way out could be to focus on a matter bounce (as white hole remnants would probably behave as pressure-less matter from the viewpoint of cosmological evolution) \cite{WilsonEwing:2012pu}. This requires a much lower-than-Planckian density at the bounce time.\\

The new scenario put forward in \cite{Bianchi:2018mml} constitutes an exciting new paradigm in black hole physics. It would be very nice to link it with the dark matter mystery but quite a lot remains to be understood.

\section{Gravitational waves}

Gravitational waves from merging black holes are now observed for real by interferometers \cite{Abbott:2016blz,Abbott:2016nmj,Abbott:2017vtc,Abbott:2017gyy,Abbott:2017oio}. This opens a new era with important interesting constraints on black hole physics and modified gravity.

\subsection{Spin in gravitational wave observations}

For a rotating black hole, the Bekenstein-Hawking entropy is given by 

\begin{equation}
S(M,j) = 2\pi M^2(1+\sqrt{1-j^2}),
\label{entr}
\end{equation}

where $j=J/M^2$ is the dimensionless spin parameter. It follows from Eq. (\ref{entr}) that, at fixed mass M, BHs with larger spin have a smaller entropy. If one assumes that  PBHs were indeed formed in the early Universe, following a microcanonical ensemble statistics,  and if we make a statistical interpretation of the BH entropy in terms of microstates, the previous statement indicates that there are fewer microstates with large spin than with small spin. In this context, the existence of a population of black holes with nearly vanishing spins is naturally predicted \cite{Eugenionew}. This is to be contrasted with astrophysical black holes, formed by the collapse of rotating stars, which are expected to be generically rotating quite fast.\\

If gravitational wave interferometers were to observe a specific distribution of events with very small spins, this would both be an evidence for the primordial origin of the considered BHs (at microcanonical equilibrium) and for the physical relevance of the Hawking-Bekenstein entropy formula.\\

To go ahead in this direction one would need to consider the entropic factor  $e^{S(m,j)}$ as the weighting of a spin distribution of PBHs determined by the physical process responsible for their creation. This distribution is however not known at this time (which means in no way that it can be approximated by a flat distribution). 

\subsection{Quasinormal Modes}

In the current LIGO/Virgo era, it would be highly desirable to make clear predictions about gravitational waves in LQG. The possibility of  detecting gravitational waves emitted by BHs before the bounce was mentioned in \cite{Barrau:2017ukm}. This could be extremely promising for opening a new window on the pre-bounce Universe thanks to the non-trivial behavior of the luminosity distance in the contracting phase, leading to a natural amplification of the signal (if the Universe is, {\it e.g.}, matter dominated). This, however, does not address the question of the specific modification to the gravitational wave shape induced by LQG corrections.\\

The best way to face this difficult question is probably to focus first on quasinormal modes (QNMs). They correspond to the ringdown phase between the transient and the exponential or power law tail in a BH merging. The radial part of the perturbed metric is described by
\begin{equation}
\Psi = A e^{-i\omega t} = A e^{-i(\omega_R + i\omega_I)t},
\label{eq:sol}
\end{equation}
where $\omega_R$ characterizes the oscillations and  $\omega_I$ the characteristic damping timescale $\tau$:
\begin{displaymath}
\tau = \frac{1}{\omega_I}\,.
\end{displaymath}
Very importantly, the frequencies of the QNMs form a countable set of discrete frequencies \cite{Chirenti:2017mwe}. 
There are actually two types of perturbations (axial and polar) in the linearized Einstein field equations described by the Regge-Wheeler and Zerilli equations. In GR, those equations are isospectral but it is not clear whether this fundamental property still holds in LQG. \\

The Regge-Wheeler equation is very close to the one used to calculate greybody factors (although the question is different : the problematics of QNM is to study the relaxation of the BH itself, not the way it scatters a quantum field). It reads for a Schwarzschild BH:
\begin{equation}
V^{\textrm{axial}}_{\ell}(r) = \left(1-\frac{2M}{r}\right)\left[\frac{\ell(\ell+1)}{r^2} - \frac{6M}{r^3}\right]\,,
\label{eq:RG}
\end{equation}
for a mode of angular momentum $\ell$. The know-how recently gained on greybody factors could therefore be usefully recycled to this purpose. It should however be clear that the technique is different (one does not search for the solution of an equation for all frequencies but for the values of the frequencies allowing for a solution with different boundary conditions) and that only models leading to substantial metric modification around the horizon might lead to observational effects. This is one of the most promising ways to relate LQG corrections to BHs with observations. 

\section{Conclusion}

The description of black holes in loop quantum gravity has much improved in the last years. A globally consistent picture is now emerging. In this article we have reviewed its possible experimental consequences.\\

The main results are the following :\\

\begin{itemize}

\item First, the Hawking evaporation spectrum should be modified in its last stages. We have shown that it could not only allow for the observation of a clear signature of LQG effects but also, in principle, to the discrimination between different LQG models. In particular, holographic models lead to specific features. The value of the Barbero-Immirzi parameter could even by measured.

\item Second, attempts to calculate the greybody factors were presented. They should keep a subtle footprint of the polymerization of space and of the existence of a non-vanishing minimum area gap.

\item Third, it was emphasized that a local quantum gravity perspective would lead to an observable modification to the Hawking spectrum (line structure), even {\it arbitrarily} far away from the Planck mass. This prediction is not washed out by the secondary emission from the BH.

\item Fourth, a model with BHs bouncing into white holes with a characteristic time proportional to $M^2$ was presented and shown to have astrophysical consequences. It can be fine-tuned to explain whether fast radio bursts or the Fermi gamma-ray excess, depending on the values of the parameters. The possible associated background was also studied. A specific redshift dependence allows to discriminate the model from other possible explanations.

\item Fifth, the possibility of having a large amount of dark matter in the form of white holes appearing after a quantum gravitational tunneling is presented together with possible weaknesses and future improvements of the model.

\item Sixth, observable effects on gravitational wave detections associated with the BHs spin distribution expected are presented.

\item Seventh, promising prospects for quasinormal modes are outlined.

\end{itemize}

It could be that black holes will play a major role in making quantum gravity become an experimental science. 

\section*{Acknowledgments}

K.M. is supported by a grant from the CFM foundation.

\bibliography{refs}
\end{document}